\documentclass[10pt]{article}

\usepackage{amsmath}
\usepackage{amssymb}
\usepackage{graphicx}
\usepackage{cite}
\usepackage{color} 

\usepackage{setspace} 
\usepackage[labelfont=bf,labelsep=period,justification=raggedright]{caption}

\makeatletter
\renewcommand{\@biblabel}[1]{\quad#1.}
\makeatother

\date{}

\begin{document}

\begin{flushleft}
{\Large
\textbf{Fractal Profit Landscape of the Stock Market}
}
\\
Andreas Gr\"onlund$^{1}$, 
Il Gu Yi$^{2}$,
Beom Jun Kim$^{2,\ast}$
\\
\bf{1} Dept. of Mathematics, Uppsala University, SE-751 06 Uppsala, Sweden
\\
\bf{2} BK21 Physics Research Division and Department of Physics,
Sungkyunkwan University, Suwon 440-746, Korea
\\
$\ast$ E-mail: beomjun@skku.edu
\end{flushleft}

\section*{Abstract}
We investigate the structure of the profit landscape obtained from the most
basic, fluctuation based, trading strategy applied for the daily stock price
data. The strategy is parameterized by only two variables, $p$ and $q$. Stocks
are sold and bought if the log return is bigger than $p$ and less than $-q$,
respectively.  Repetition of this simple strategy for a long time gives the
profit defined in the underlying two-dimensional parameter space of $p$ and
$q$. It is revealed that the local maxima in the profit landscape are spread in
the form of a fractal structure.  The fractal structure implies that successful
strategies are not localized to any region of the profit landscape and are
neither spaced evenly throughout the profit landscape, which makes the
optimization notoriously hard and hypersensitive for partial or limited
information.  The concrete implication of this property is demonstrated by
showing that optimization of one stock for future values or other stocks
renders worse profit than a strategy that ignores fluctuations, i.e., a
long-term buy-and-hold strategy.

\section*{Introduction}
Everyone wants to be rich, who doesn't? As a way of investment,
stock market provides not only a chance to become a millionaire, but
also a direct shortcut to pennilessness. In the present paper we focus on two important properties
of the stock trading industry. First, it is shown that individual households pay a tremendous
performance penalty for active trading~\cite{barber}.
Even professional fund managers cannot outperform the market indices in long term~\cite{malkiel}.
Second, many stock traders can be characterized as chartists, in principle using stock charts solely to make trading decisions~\cite{lo,menkhoff}.
Inspired by the above two observations, we will try to make some conclusions regarding the chances of
actually beating the market trend by exploiting temporal variations in the stock market.
We will not try to perform a complete modeling of interacting traders, nor will we model all possible trading strategies, but rather learn from a very simple strategy that exploits the fluctuations of stock prices and see whether it can shed any light to the observed difficulties in beating the market trend.

We admit that the trading strategies used in reality could be much more
complicated and sophisticated than the naive strategy in this work.  
Various
technical trading strategies based on moving averages, price momentum, 
channel breaking, and relative stock index have been being 
used~\cite{schulmeister,sullivan,park}. In
technical trading, suitably defined signals of buying and selling 
are produced from the past stock price movements, and trading
strategies can typically be grouped into two different categories:
trend-following and contrarian.  The stock trading strategy in this work is
parameterized by only two parameters that will be used for quantifying how
fluctuations propagate to the long-term profit of the strategy.  It should be
noted that we are not aiming at proposing a profitable trading strategy, but we
hope to understand the structure of the profit landscape yielded from a very
basic trading strategy of buying and selling.  However, by composing a very
simple strategy from elementary buying and selling signals, more complex
strategies can be composed from combinations of such signals, and robust
features of our strategy, revealed by the profit landscape, should hold also
for such, more complex, strategies.

\section*{Methods}
In this work, we use 95 US company stocks ($i=1,2, \cdots, 95$), which existed
for 21 years between 1983 and 2004.  The stock price
$x_i(t)$ is given for the $i$-th stock at time $t (t=1,2,\cdots,T = 5301)$,
which is the consecutive integer increasing by one at every trading date.
What we mean by a strategy $S$ in this paper is the way of determining (i)
whether or not to trade (buy or sell), and if it is decided to trade (ii) how
many units of stocks are to be traded.  Only for simplicity, we assume that if
`buy' decision is made in (i), we buy number of stocks
by spending the $f_b$ fraction of cash.  Likewise, if `sell' decision is made,
$f_s$ fraction of stocks in possession are sold. Accordingly,
our imaginary simple portfolio (we trade stocks for only one
company, i.e., $i$ fixed) is composed of only two accounts; cash $m(t)$ and
number $n(t)$ of stocks, and the estimated value of the portfolio is given as
the sum of the cash and stocks, $m(t) + n(t) x(t)$. In reality,
the amount of cash alone can increase in time by trading the
risk-free asset, which is not taken into account in the present study
only for the sake of simplicity. In other words, we assume that
the risk-free interest rate is zero. In the same spirit, we also neglect
the increase of cash via dividends provided by companies.
We note that our expression for the value of the portfolio is the same as
that of the wealth in Ref.~\cite{yeung}.
In order to quantify a strategy with a small set of parameters,
we propose the following simple strategy (we call it $S_1$):
\begin{itemize}
\item The trade decision at time $t$ depends only on the stock prices
at two distinct times $t$ and $t' (<t)$. More specifically, the log return
$R(t,t') \equiv \ln [x(t)/x(t')]$ is used for the trade decision.
\item If $R(t,t') > p$, one judges that the stock price went up much and thus
decides to sell.
\item If $R(t,t') < -q$, the decreased stock price makes the stock attractive
to buy, and thus one decides to buy.
\end{itemize}

We also consider the inverse strategy called $S_2$:
\begin{itemize}
\item If $R(t,t') > p$, one  expects that the stock price will go up later and decides to buy.
\item If $R(t,t') < -q$, one is afraid of further price falling, and thus decides to sell.
\end{itemize}

In words, the strategy $S_1$ can be termed as ``sell-on-rise/buy-on-fall'', whereas
the inverse strategy $S_2$ as ``sell-on-fall/buy-on-rise''. Similarly, the
trend-opposing (i.e., contrarian) and the trend-following strategies have been used in
Ref.~\cite{yongjoo}, although the meaningfulness of the concept of ``trend''
in real markets is questionable~\cite{malkiel}. For the sake of simplicity, we
impose the non-negative cash constraint ($m(t) \ge 0$)~\cite{yongjoo} and
the non-negative stock constraint ($n(t) \ge 0$).
In other words, you can neither buy stocks if you do not have enough cash,
nor you can sell stocks which you do not have (short selling is not allowed).
We also assume that the stock price is given exogenously and that
the transaction price of the stock trading equals the daily closure price in the
data file.

In general, the strategy in the present setup can be written as
$S_{1,2}(p, q; f_b, f_s, d)$ with $t' = t - d$ ($d \geq 1$).
For given values of $f_b$, $f_s$, and $d$, we investigate the performance
of the investment
strategy parameterized by the two variables $p$ and $q$.
The detailed algorithm of performing our investment game is as follows:
(i) We pick a company $i$ one by one. (ii) At the first trading date, we start from
one million dollars $m(0) = 10^6$.
(iii) For given values of $p$ and $q$, we keep
applying the strategy repeatedly till the last date $T$. (iv) We evaluate the
performance of strategy by computing the profit defined by $\Pi(p,q) \equiv [m(T) + n(T) x(T) - m(0)]/m(0)]$.
For each stock trading, we also assume that we pay small trading fee $0.1\%$.
In the real stock
market, there often exits minimum number of stocks to be traded. We have compared 1, 10, and 100
as minimum trading volume, only to find insignificant change of results. Henceforth, our
strategies allow the trade of a single unit of stocks (but you cannot sell or buy a fraction of a stock).

\section*{Results}
\begin{figure}
\includegraphics[width=0.98\textwidth]{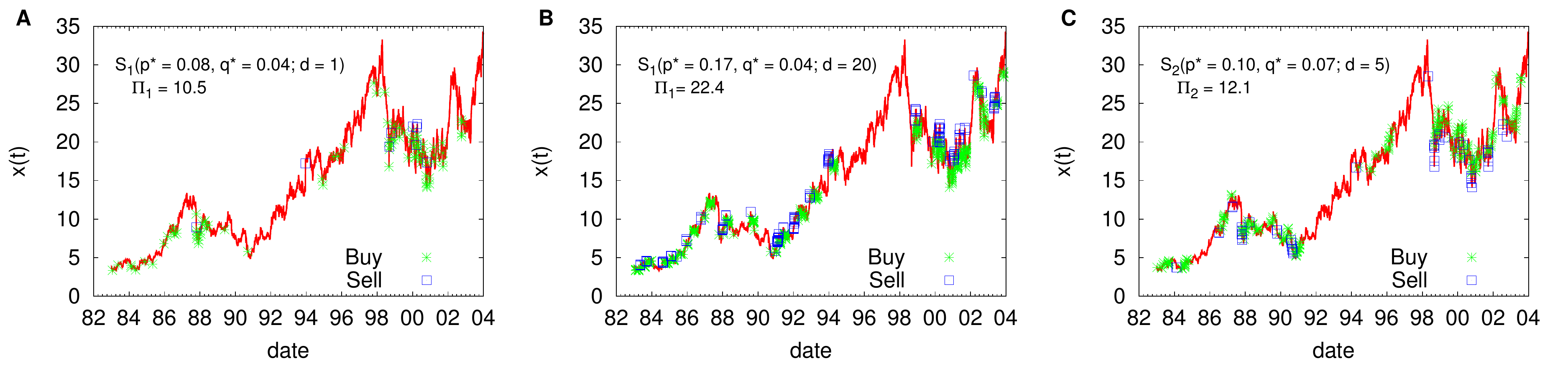}
\caption{ {\bf Price time series and strategy-dependent trading decisions.}
The time series of the stock price for a company is plotted as a
function of time (full line), and the time instants when buy and sell decisions
are made for given strategies are marked as star and square symbols, respectively.  For the
resolution $N\times N$ ($N=1024$) in the parameter space of $(p,q) \in
[0,1]\times[0,1]$, we compute the profit $\Pi(p,q)$ at $N^2$ points to find the
global profit maximum at $(p^*, q^*)$.  The used strategies are
(A) $S_1(p^*=0.08, q^* = 0.04; d=1)$, (B)
$S_1(p^*=0.17, q^*=0.04; d=20)$, and (C) $S_2(p^*=0.10, q^*=0.07; d=5)$.
For all cases, we used $f_b = f_s = 1/2$.
Depending on the strategy used, the instants when buy and sell decisions are made
are very different from each other, and the resulting profit values are also quite
different: $10.5$, $22.4$, and $12.1$ in (A), (B), and (C), respectively.
}
\label{fig:xttrade}
\end{figure}

We first compute the profit $\Pi^i(p,q)$ of historic stock data for the $i$-th company,
in the two-dimensional parameter space $(p,q)$ defined on unit square $[0,1]\times[0,1]$.
For convenience, we discretize the parameter space into an $N \times N$ square grid and compute
$N^2$ values of $\Pi^i(p,q)$ at the center of each small square of the size $(1/N)\times(1/N)$.
Once those values are computed, one can easily  find the global maximum of the profit $\max_{(p,q)} \Pi^i (p,q)$
for the given resolution $N$, and call those optimal values as $(p_i^*, q_i^*)$.
We  also calculate the number $M$ of local maxima which satisfy the
criterion that four neighboring points in square grid have smaller values of the profit than at the center.
As an example, we run the simulations at $N^2$ ($N=1024$) grid points
for $f_b = f_s = 1/2$  and then use the obtained optimal value $(p^*_i, q^*_i)$ to construct
Fig.~\ref{fig:xttrade}, where we also denote time instants when the buy and the sell decisions are made.

Whether or not short term fluctuations in the stock market may be exploited is addressed by benchmarking
our strategies with a ``buy-and-hold'' strategy $S_0$ (see Table ~\ref{tab:strategy}):
At $t=1$ all cash is spent to buy stocks and the profit is evaluated at $t=T$,
given by $\Pi_0 = [x(T) - x(1)]/x(1)$.
The value of $\Pi_0$ changes from company to company, depending on the
growth rate of each firm. The growth rate conditioned to the firm size
has been known to have exponential distribution function~\cite{growthrate},
while the size of each firm is power-law distributed~\cite{axtell}.
For a company shown in Fig.~\ref{fig:xttrade}, $\Pi_0 = 8.76$,
which is smaller than the maximum profit realizable by the strategies
$S_1$ and $S_2$.
The actual profit of any strategy for historic data is however of marginal interest; more interesting issue to pursue is what we can learn from the profit landscape and see if it allows us to make money in the future.

\begin{figure}
\includegraphics[width=0.38\textwidth]{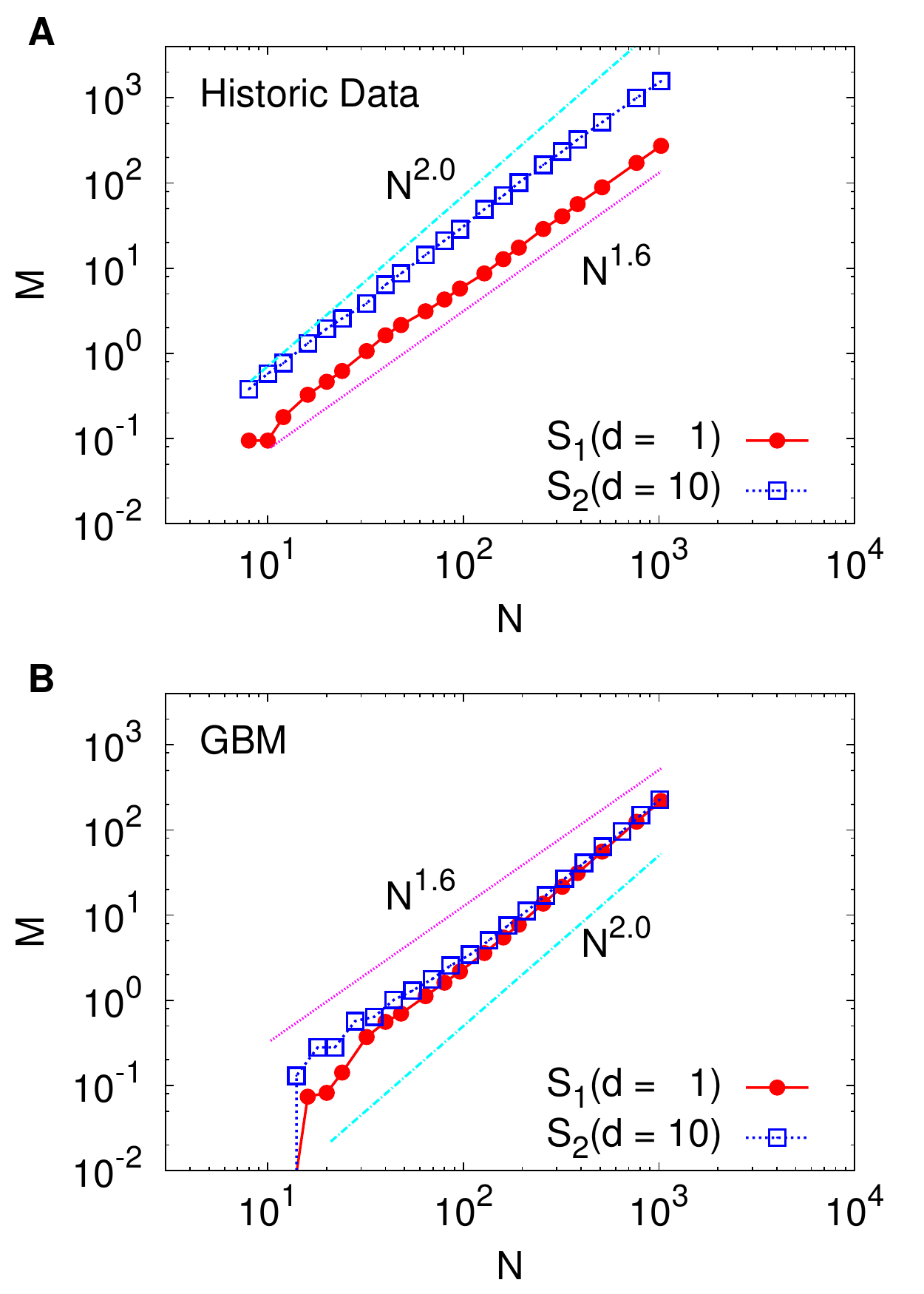}
\caption{{\bf Scaling of the number of local maxima in the profit landscape.}
(A) The number $M$ of local maxima of the profit
landscape [$\Pi(p,q)$] is computed as a function of the number $N$ of grid
points in the two-dimensional parameter space. The local maxima of the profit
are distributed like a geometric fractal, manifested by $M \sim N^a$ with $a
\approx 1.6$. For comparison, we also display the curve for $M \sim N^2$,
which clearly deviates from the actual result.
(B) $M$ versus $N$ calculated from the GBM time series (see text).
For GBM, $M \sim N^2$ fits better to the result than $M \sim N^{1.6}$.
}
\label{fig:fractal}
\end{figure}

We next study how the number $M$ of local maxima of the profit landscape defined
on the $p$-$q$ plane changes as the resolution parameter $N$ is increased.
If the profit landscape has a simple structure that there are only a few
number of peaks, $M$ is expected to first increase with $N$, since
the number of peaks found in the higher resolution (larger $N$) will be larger than
that in the lower resolution (smaller $N$). In this case of a simple landscape, $M$
will then soon saturate to an $O(N^0)$ value, and will not increase any more even when $N$ is
increased further. If this is indeed the case, the optimization process
we described above can be very efficient to locate the global maximum
profit. 
If the maxima are uniformly distributed they will scale linearly with the
number of boxes $N\times N$, that is $M\sim N^2$. 
In Fig.~\ref{fig:fractal}A we show how $M$ (averaged over 95 USA stocks)
changes as a function of $N$ for different strategies.  The scaling falls
between the two extremes: Surprisingly, $M$ increases with $N$ following an
algebraic form of $M \sim N^{1.6}$ for both strategies.
Although not reported here, we find that
the behavior $M \sim N^a$ with $a \approx 1.6$ for broad range of parameter
values of $f_b, f_s$, and $d$, both for $S_1$ and $S_2$. 
The same value of $a$ is found when extended range 
$(p,q) \in [0,2] \times [0,2]$ is used, and when eight neighbors of square grid
(instead of four nearest neighbors) are compared for the determination of local maxima.  Furthermore,
Korean stock price data also reveal the same behavior with $a \approx 1.6$ (to
be reported elsewhere).
These observations clearly indicate that the pattern of how local maxima are
scattered in the two-dimensional parameter space is described by a {\it
fractal} structure.

We next compare our findings to the standard model of stock price, i.e., geometric
Brownian motion (GBM)~\cite{black1973pricing}:
\begin{equation}
\label{eq:gbm}
dX_t=\mu X_t dt + \sigma X_t dW_t\, ,
\end{equation}
where $X_t$ is the Brownian random variable and $W_t$ is the Wiener process.
The parameters $\mu$ and $\sigma$ are fitted to
each of the stocks and a number of replicas are simulated over the same period
of the real stock data. A striking difference to the real stock price data is
that for the GBM local maxima are evenly distributed, seen by the scaling
$M\sim N^2$ in Fig.~\ref{fig:fractal}B. A conclusion we can make from this is
that the profit cannot be explained solely from the first and the second order
statistical moments (the mean value and variance) but a more detailed
description of stock price movement is needed. 
It is well known
that GBM is not able to describe some stylized facts of real price movements such as high value
of kurtosis, fat tails in the probability distributions of log-returns, and the stochastic
volatility and its clustering behavior~\cite{stanleybook,schoutensbook,fusai,mand63,Yamasaki28062005}. 
Beyond GBM various stochastic processes
have been proposed with  L\'evy processes as the most prominent 
example. Various models as subclasses of L\'evy processes and autoregressive models have been 
proposed~\cite{stanleybook,schoutensbook,fusai,cowpertwait}.
The results from more sophisticated  models of financial-time series are 
however not further corroborated here.

Although all the strategies belonging to $S_1$ and $S_2$ have
qualitatively the same fractal structure in their profit landscapes, the
maximum profit realizable by a given strategy is very different from each
other.  In general, the average maximum profits of $S_1$ are found to be
larger than those of $S_2$.
This can also be seen in
Fig.~\ref{fig:xttrade}B and C: In $S_1$ sell prices are often higher
than the buy price, while the opposite is seen for $S_2$.  
Similarly, contrarian strategies have been shown to yield higher profits than
trend-following ones in \cite{schulmeister}.

We also observe
that for given values of $f_b$ and $f_s$, the average maximum profit does not
significantly depend on the value of $d$. This is not a surprising observation
since it is well known that the autocorrelation of returns decays very fast (in
several minutes)~\cite{4min}, and thus the time difference $d$ larger than one
day will not make much difference in the results.

We have seen that there exist fluctuation based strategies for which the
maximum profits significantly exceed the profit from the long term buy-and-hold
strategy $S_0$.  The question is how to find the optimal strategy,
since we need to estimate $p$ and $q$ for future stock data by
using past stock data. Even if we possibly cannot find the \emph{optimal}
strategy by optimizing for old data we would like to know whether we can be sure
to find a \emph{good} strategy from such an optimization process. The fractal
profit landscape suggests that the optimization process is very sensitive to
perturbations, such as missing data and the change of optimization period of time.

We study the stability of the optimized strategies in two different ways:
First, we compare how the profit changes across different
companies for given values of $p$ and $q$. More specifically, we obtain the
optimized values $p^*_i$ and $q^*_i$ for each company $i$, and use the average
values $\langle p^* \rangle \equiv (1/95)\sum_{i=1}^{95} p^*_i$ and $\langle q^*
\rangle \equiv (1/95)\sum_{i=1}^{95} q^*_i$ to compute the profit by the strategy
$S_1( \langle p^* \rangle,  \langle q^* \rangle; f_b = f_s = 1/2, d = 1)$
applied for each company $i$, which gives us the average profit $\langle
\Pi_1(\langle p^* \rangle,  \langle q^* \rangle) \rangle$.  Second, we compare how
the profit changes in different time periods as follows: We use the first half
period ($t=1, \cdots, T/2$) for optimization of the strategy, and then use the
obtained values $p^*_i$ and $q^*_i$ to compute the profit for the second half
period ($t=T/2+1, \cdots, T$). These two ways to test the stability of the
optimized strategy can be phrased as the tests for spatial stability (across
stocks $i$) and for temporal stability (across the time $t$).

\begin{figure}
\includegraphics[width=0.40\textwidth]{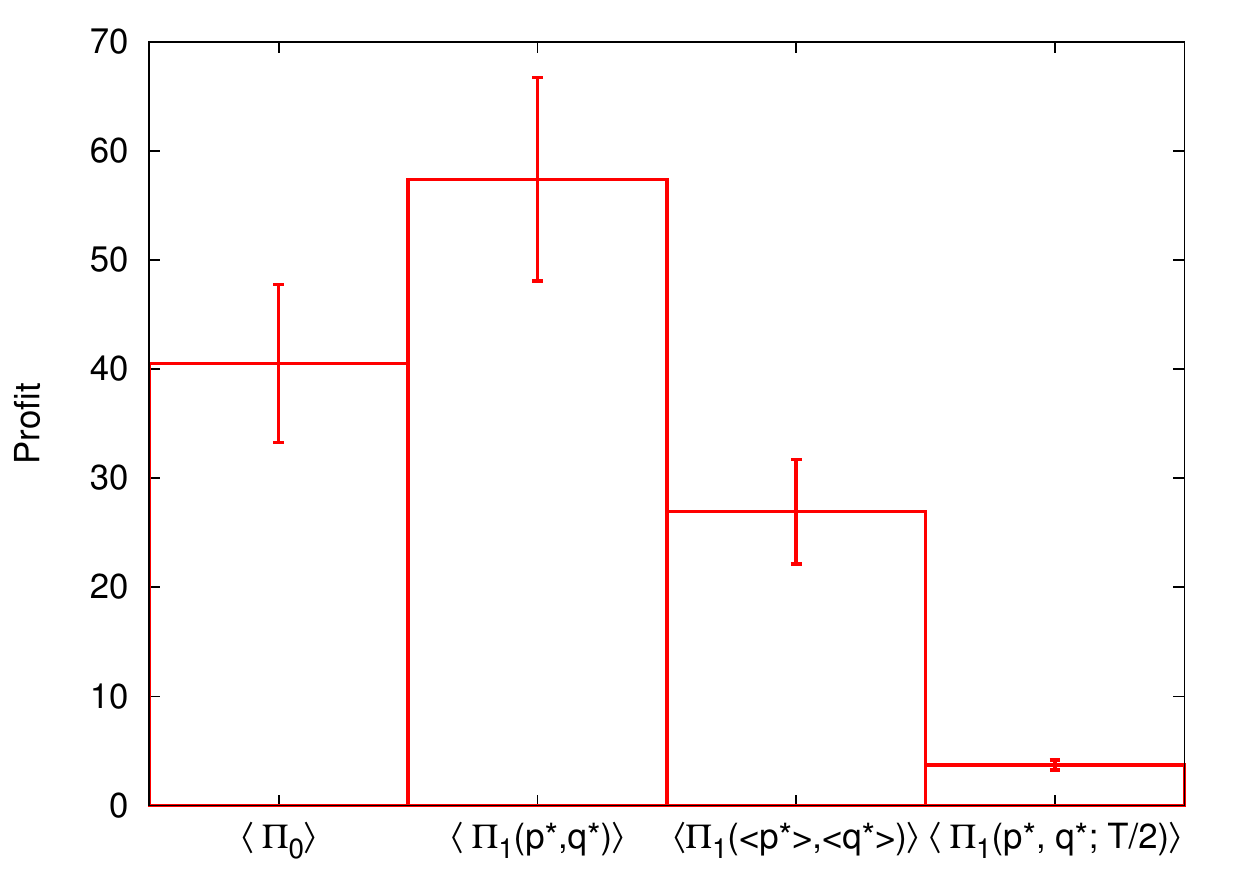}
\caption{{\bf Performances of different strategies.}
$\Pi_{0,1,2}$ is the profit from the strategy $S_{0,1,2}$
(see Table~\ref{tab:strategy}), and $\langle \cdots \rangle$ is the
average over all companies. See text for details.
}

\label{fig:performance}
\end{figure}

Figure~\ref{fig:performance} summarizes the results from the stability tests of
the optimized strategy.  It is again shown that the optimized strategy for each
individual company for the whole period yields better profit than that from the
buy-and-hold strategy, i.e., $\langle \Pi_1(p^*,q^*) \rangle > \langle
\Pi_0 \rangle$.  Since each stock has
different time evolution behavior, the optimized values $p^*_i$ and $q^*_i$ are
different for different companies. In this regard, it makes some sense to
use $\langle p^* \rangle$ and  $\langle q^* \rangle$ for
all the companies, expecting that these values could have somehow better
performance, although not as good as $\langle \Pi_1(p^*_i,q^*_i) \rangle$.
The third box denoted as $\langle \Pi_1 (\langle p^* \rangle, \langle q^*
\rangle) \rangle$  in Fig.~\ref{fig:performance} clearly shows that the use of
the average values $\langle p^* \rangle$ and $\langle q^* \rangle$ does not
give us better profit than the buy-and-hold strategy. In
Fig.~\ref{fig:performance}, we also display the result from the temporal
stability test, denoted as $\langle \Pi_1( p^*, q^*; T/2) \rangle$.
It shows that the use of the optimized strategy for later time periods
dramatically reduces the profit value. 
The same conclusion is reached when we use finer time windows as follows: We
divide the whole time period $T$ into 20 time intervals and obtain the
optimized value $p_i^*(\tau)$ and $q_i^*(\tau)$ at each $\tau$-th interval
($\tau = 1,2,\cdots,20$). We then use $p_i^*(\tau)$ and $q_i^*(\tau)$ to
compute $\Pi_1(\tau')$ at later time interval $\tau'$ ($\tau' > \tau$). Even
the largest value of the profit [$\max_{\tau' > \tau} \Pi_1(\tau')$] is found
to be smaller than the profit from the buy-and-hold strategy at the same
interval. This indicates that our conclusion of the temporal instability of the
optimal strategy holds at least for sufficiently long time scales.

The results from the stability tests can be summarized as follows:
Even though fluctuations, theoretically, may be exploited in our basic trading scheme,
the $p,q$-parameters (i) cannot be estimated from historic data and (ii) nor can they be estimated from other stocks.
In both cases the performance is significantly worse than the buy-and-hold strategy.
Our results agree with other existing studies: It is now generally believed that
stock market has become more efficient and most technical trading strategies based on daily price
changes stopped being profitable in mature market~\cite{schulmeister}, although developing
markets can still be different~\cite{park}.

\begin{figure}
\includegraphics[width=0.90\textwidth]{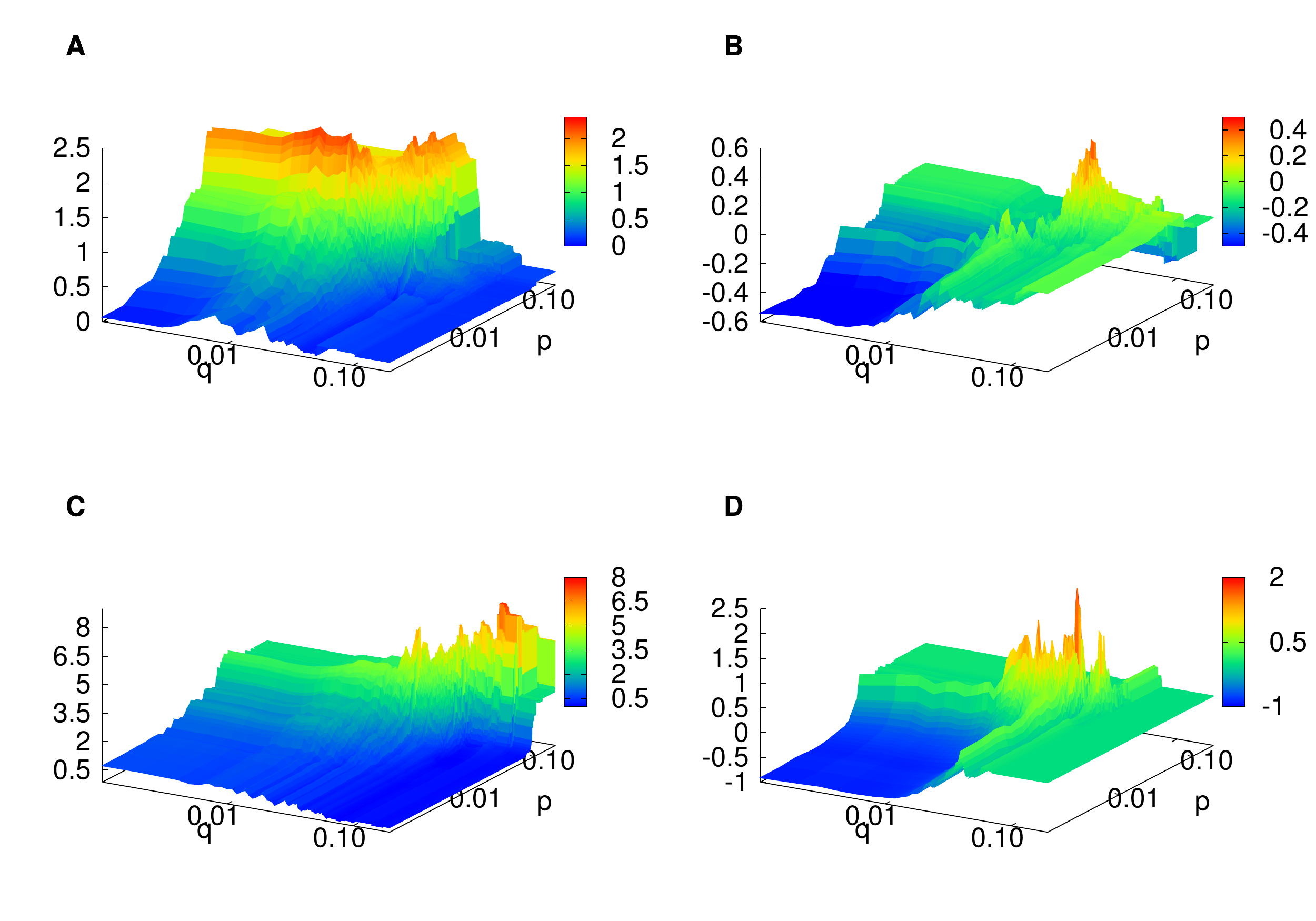}
\caption{{\bf Profit landscape in two-dimensional parameter space.} 
(A) $\Pi_1^i(p,q; t \in [1,T/2])$ and (B) $\Pi_1^i(p,q; t \in [T/2+1,T])$ 
are for the same stock ($i$) but for different time periods, and
(C) $\Pi_1^j(p,q; t \in [1,T/2])$ is for other stock ($j \ne i$).
The three landscapes (A)-(C) look different, in agreement with 
the instability of optimal value of $p$ and $q$ over different stocks
and different time periods. (D) $\Pi_1(p,q; t \in [1,T/2])$ for a
GBM time series. For better visibility in the small values of $p$ and $q$, 
we plot the landscapes in the plane of $p$ and $q$ in log scales.
For (A)-(D), the strategy $S_1$ is used.
}
\label{fig:landscape}
\end{figure}

We next investigate the actual shapes of the profit landscapes in terms of the observations
made above. Figure~\ref{fig:landscape} displays (A) $\Pi_1^i(p,q;t\in[1,T/2])$ and 
(B) $\Pi_1^i(p,q;t\in[T/2+1,T])$ for one stock ($i$), and (C)  $\Pi_1^j(p,q;t\in[1,T/2])$
for other stock ($j$). One can see that the shape of the landscape looks quite different
from each other, supporting the above conclusion of the instability of optimal strategy.
It is also seen that the profit landscape is quite rugged in accordance with the fractal-like
distribution of the local maxima reported in the present study.
For comparison, we also show in Fig.~\ref{fig:landscape}D the profit landscape for the GBM
series. The difference of the fractal dimension for local maxima between
real stock prices and GBM is not clearly seen in Fig.~\ref{fig:landscape}, however,
the landscape for (D) GBM looks more rugged than the landscape for (A) an actual stock.

\section*{Discussion}
In summary, we have investigated the profit landscape defined by a set of
simple investment strategies. We have shown that the local maxima in the profit
landscape are scattered like a fractal and that the global profit maximum
increases very slowly as the search resolution in the parameter space is
increased. These findings imply that a local search in a strategy space to get
the highest profit is almost impossible.
We believe that our conclusions of poor performances of fluctuation-based
trading strategies and their fractal profit landscapes are related with the unavoidable
lack of future information of a company. If one has full information of the stock price change
for the future, that information can directly be used to optimize trading strategy.
However, if the future information is not sufficiently accurate, it can be basically
useless in increasing profit, as has already been shown in Ref.~\cite{toth}.
We plan to study the rugged profit landscape in the present work
in comparison with the fitness landscapes in other research areas~\cite{weinberger}.

\section*{Acknowledgments}
This work  was supported by the National Research Foundation of Korea(NRF) grant
funded by the Korea government(MEST) (2011-0015731).


\section*{Figure Legends}

\section*{Tables}

\begin{table}
\caption{ {\bf Strategy set.} For $S_{1,2}$, the buy/sell decisions are made
according to the log-returns of the stock price with the time difference $d$.}
\label{tab:strategy}
\begin{tabular}{ccc}
\hline\hline
strategy  & description \\
\hline
$S_0$ & buy-and-hold \\
$S_1$ & sell-on-rise and buy-on-fall  \\
$S_2$ & buy-on-rise and sell-on-fall  \\
\hline
\end{tabular}
\end{table}

\end{document}